\documentstyle[twoside,fleqn,espcrc2,epsfig]{article}
\newcommand{\AmS}{{\protect\the\textfont2
  A\kern-.1667em\lower.5ex\hbox{M}\kern-.125emS}}

\title{Non-resonant diagrams in radiative four-fermion processes}

\author{ J. Fujimoto, T. Ishikawa, S. Kawabata, Y. Kurihara, Y. Shimizu
\address{ Minami-Tateya Collaboration, KEK, Japan }
        and
        D. Perret-Gallix\address{LAPP-IN2P3/CNRS, France} }

\begin{document}

\begin{abstract}
The complete tree level cross section for
$e^+e^- \to e^- \bar\nu_e u \bar{d} \gamma$ is computed and discussed
in comparison
with the cross sections for $e^+e^- \to e^- \bar\nu_e u \bar{d} $ and
$e^+e^- \to \bar{u} d u \bar{d}$.
Event generators based on the GRACE package for the non-radiative
and radiative case are presented. Special interest is brought
to the effect of the non-resonant diagrams overlooked so far in other
studies. Their contribution to the total cross section
is presented for the LEP II energy range and
for future linear colliders ($\sqrt s$ =500 GeV).
Effects, at the W pair threshold, of order 3\% ($e^- \bar\nu_e u \bar{d}$)
and 27\% ($\bar{u} d u \bar{d}$) are reported.
Similar behaviour for the radiative case is shown.
At $\sqrt s$ = 500 GeV, the relative contribution of the non-resonant
diagrams for the radiative channel reaches 42.5\%.
\end{abstract}
\maketitle
\section{Introduction}
Two years ago, a complete calculation of two four-fermion final
states in $e^+e^-$ collisions, $e^+e^- \to e^-\bar\nu_e u \bar{d}$ and
$e^+e^- \to \bar{u} d u \bar{d}$, was reported at the Sotchi
meeting \cite{sotchi}. It was shown that the non-resonant
diagrams play an important role below the W pair threshold (up to 27\%
at $\sqrt s =$ 150 GeV) and exhibit a non-negligible effect at higher
energy (7\% at $\sqrt{s}=$ 190 GeV). In these results, in addition to
a set of experimental loose constraints, a cut on the $\theta$ angle
of the outgoing
electron was applied to cope with gauge violations appearing in the
subset of $\gamma -W$ diagrams. Below threshold, the non-resonant
diagrams competing only with the off-shellness of the resonant ones,
their relative contribution
becomes quite large. The non-resonant diagrams
are dominated by $t$-channel $\gamma-W$ graphs in similarity with the
well-known two-photon ($\gamma-\gamma$) processes.
Hence the cross section increases
with the energy and the contribution of non-resonant processes
keeps growing in contrast to that of resonant diagrams which
falls with the energy. This effect was confirmed at this workshop
by the analysis of selected final states based on
a dedicated four-fermion generator \cite{pittau}.
The non-resonant processes are essentially single W or no W production.
Some examples of these type of diagrams for the radiative case
can be seen on Fig.1b,2.
As far as the W mass reconstruction is concerned, these diagrams should
be counted as background processes, and they will contribute to
the overall mass and width measurement uncertainties.

It has been recognized, since some time \cite{veltman},
that the radiative corrections
in $W$ pair production
have to be included for a precise comparison with measurements
at the $e^+ e^-$ colliders.
\begin{figure}[ht]
\begin{center}
\vspace*{-0.5cm}
%\hspace*{-2.cm}
\mbox{\epsfig{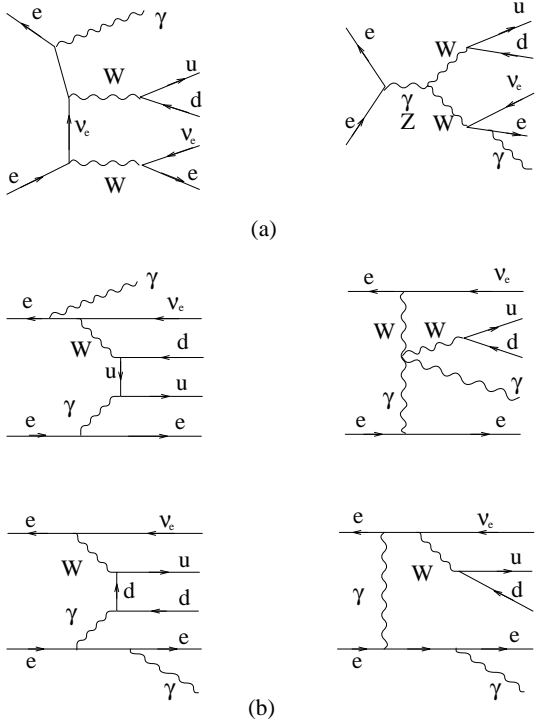}}
\vspace*{-0.5cm}
\caption{Some examples of diagrams from the radiative process
 $e^+e^- \to e^- \bar\nu_e u \bar{d}\gamma$: a) Resonant diagrams,
b)Non-resonant $\gamma-W$ diagrams.}
\end{center}
\vspace*{-1cm}
\end{figure}
\begin{figure}[ht]
\begin{center}
\vspace*{-0.5cm}
%\hspace*{-1.8cm}
\mbox{\epsfig{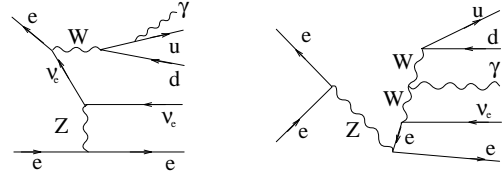}}
%\vspace*{-8.5cm}
\vspace*{-0.5cm}
\caption{(Fig.1 cont.) Some examples of diagrams from the radiative process
$e^+e^- \to e^- \bar\nu_e u \bar{d}\gamma$: Other non-resonant diagrams.}
\end{center}
\vspace*{-1cm}
\end{figure}
Initial state radiations (ISR) reduce the available center of mass
energy. Close to the W pair threshold, ISR will, therefore, enhance the
relative
contribution of non-resonant diagrams. Final state radiative corrections
will contribute to additional uncertainties on the mass reconstruction,
unless the photon is detected and properly accounted for in the mass
reconstruction algorithm. However, photons coming from the
intermediate W, which can be mistakingly recognized as final state
bremsstrahlung, will introduce additional uncertainties.

Although a number of studies has been carried out in these last years,
including
\cite{shimizu}\cite{other},
we present, for the first time, the complete tree level
computation of one of the radiative four-fermion processes.
This work can be used as a benchmark for testing the validity of
simplified and therefore faster generators.
\section{The GRACE system}
The GRACE system \cite{grace}
has been used to perform the very lengthy computations involved in the
study of most radiative processes. The GRACE package is a complete
set of tools for computing
tree level processes. All the phases involved in a given computation
are covered: from the process definition (specification of the initial
and final particles) to the event generator. It is composed essentially
of three components: the diagram generator, the construction of
the matrix element based on helicity amplitude function from the
CHANEL \cite{chanel} library and the multi-dimensional phase space integration
package BASES
\cite{bases} associated with the event generator
SPRING \cite{bases}.

Fermion masses are properly introduced in the helicity amplitudes.
The particle width is introduced into the gauge boson propagators when
the denominator can vanishes for positive squared momentum transfer.
A gauge invariance checking program is automatically built by
the system.

Only the kinematics for complex processes has to be written by
the user. The issue, here, is to figure out what is the best transform
to apply to the basic integration variables to regularize
the divergencies appearing in the integration of the matrix element.
The best kinematics will give the best accuracy for a given computation
time.
\section{Four-fermion production}
The results presented hereafter have been obtained using the following
set of parameters:
\begin{eqnarray*}
M_Z & = & 91.1~{\rm GeV}\\
\Gamma_Z & = & 2.534~{\rm GeV}\\
\alpha & = & 1/137\\
\sin^2\theta_W & = & 1-(M^2_W/M^2_Z)\\
M_W & = & 80 ~{\rm GeV}\\
m_u &=&m_d~=~0.1~{\rm GeV}.
\end{eqnarray*}
The width of the W is taken from the Particle Data Table;
$\Gamma_W=2.25$ GeV. The gauge boson (W,Z)
widths are assumed to be constant in the calculation. Furthermore some
realistic experimental cuts have been introduced:
\begin{eqnarray*}
\theta_e & > & 8,~20,~30,~40^\circ\\
172^\circ & > & \theta_\gamma,~ \theta_{q,\bar q} > 8^\circ\\
E_\gamma & > & 1 ~{\rm GeV},~ E_{e,q,\bar q} > 1 ~{\rm GeV} \\
\end{eqnarray*}
where $\theta_e,\theta_\gamma$ and $\theta_{q,\bar q}$ are angles
measured from the incident $e^-$ beam and $E_{e,q,\bar q}$ are
the energies of final $e^-,q$ and $\bar q$, respectively.
The electron polar angle cut is introduced to avoid an undesirable
gauge violation for extremely forward directions \cite{boos} caused inevitably
by the finite width of W boson when the subset of $\gamma -W$ diagrams
is considered.

The gauge invariance of the amplitude is checked numerically for
a random selection of the boson gauge parameters at several points
in the phase space. The errors are within the precision of numerical
calculation (typically less than $O(10^{-12})$ in double precision) when
the W width is turned off. Further gauge cancellations are checked to
confirm that the obtained cross sections are stable against finite
width. The procedure is as follows: first extract the product of gauge boson
propagators without width as an overall factor of the whole amplitude.
Then replace them by those with finite width. As a result, one is left
with rigorously gauge invariant amplitude\cite{pukhov}. For the resonant
diagrams,
this method gives the same results as ordinally method. While for the $\gamma
-W$
diagrams, this gives 10\% smaller results.
\begin{figure}[ht]
%\begin{center
\vspace*{-1.5cm}
%\hspace*{-2.cm}
\mbox{\epsfig{file=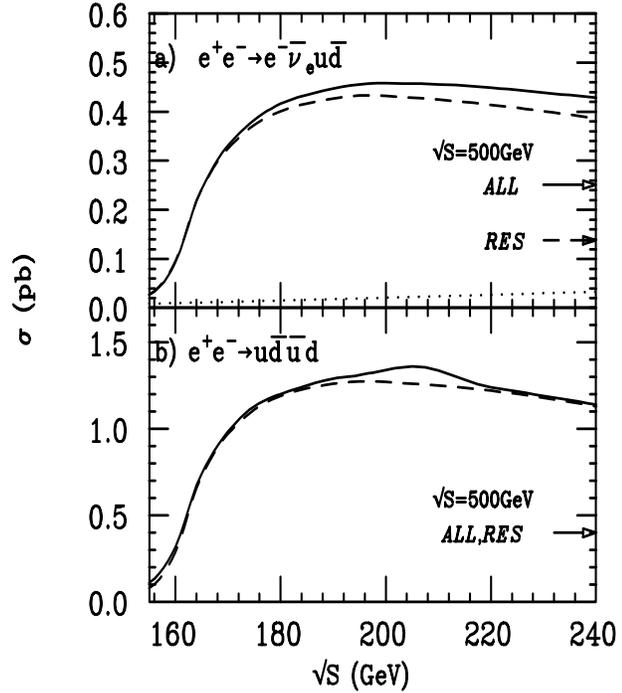,width=9.5cm,height=10.cm}}
\vspace*{-1cm}
\caption{Cross sections for a) $e^+e^- \to e^- \bar\nu_e u \bar{d}$
and b) $e^+e^- \to \bar{u} d  u \bar{d}$ versus C.M. energy
with $\theta_e \ge 8^\circ$.
The solid curves are for all
diagrams and dashed ones for three resonant graphs. The dotted line in a) is
the contribution from the $\gamma-W$ and $Z-W$ processes.
The estimates at 500 GeV are indicated by arrows.}
\vspace*{-0.5cm}
%\end{center}
\end{figure}
\begin{figure}[ht]
%\begin{center
\vspace*{-1cm}
%\hspace*{-2.cm}
\mbox{\epsfig{file=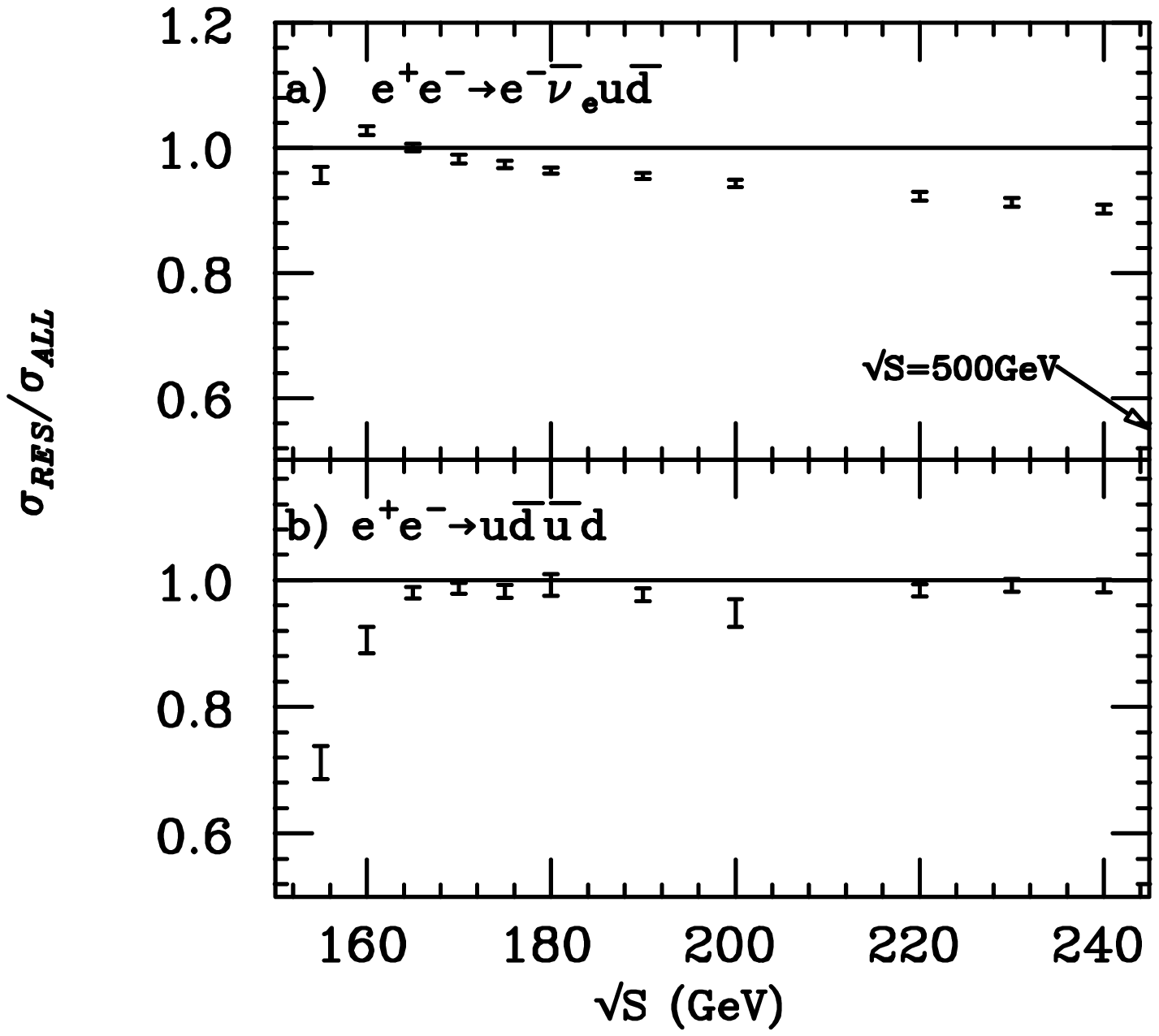,width=9.5cm,height=10.cm}}
\vspace*{-1cm}
\caption{Energy dependence of the relative contribution of resonant diagrams
in a) $e^+e^- \to e^-\bar\nu_e u \bar{d}$ and b)
$e^+e^- \to \bar{u} d  u \bar{d}$ with $\theta_e \ge 8^\circ$
to the full computation.
}
\vspace*{-0.5cm}
%\end{center}
\end{figure}
\begin{figure}[ht]
\vspace*{-1cm}
%\hspace*{-2.cm}
\mbox{\epsfig{file=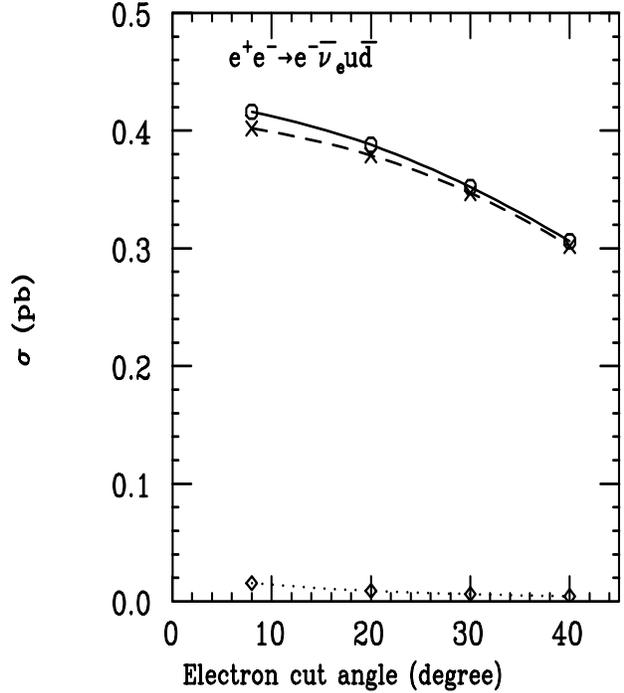,width=9.5cm,height=10.cm}}
\vspace*{-1cm}
\caption{
Electron angular cut dependence for
$e^+ e^- \to e^- \bar \nu_e u \bar d$
(C.M. energy = 180 GeV). Solid line: all diagrams, dashed line: three resonant
diagrams and dotted line: $\gamma-W$ and $Z-W$ processes.
}
\vspace*{-0.5cm}
\end{figure}
\begin{figure}[ht]
%\begin{center
\vspace*{-1cm}
\mbox{\epsfig{file=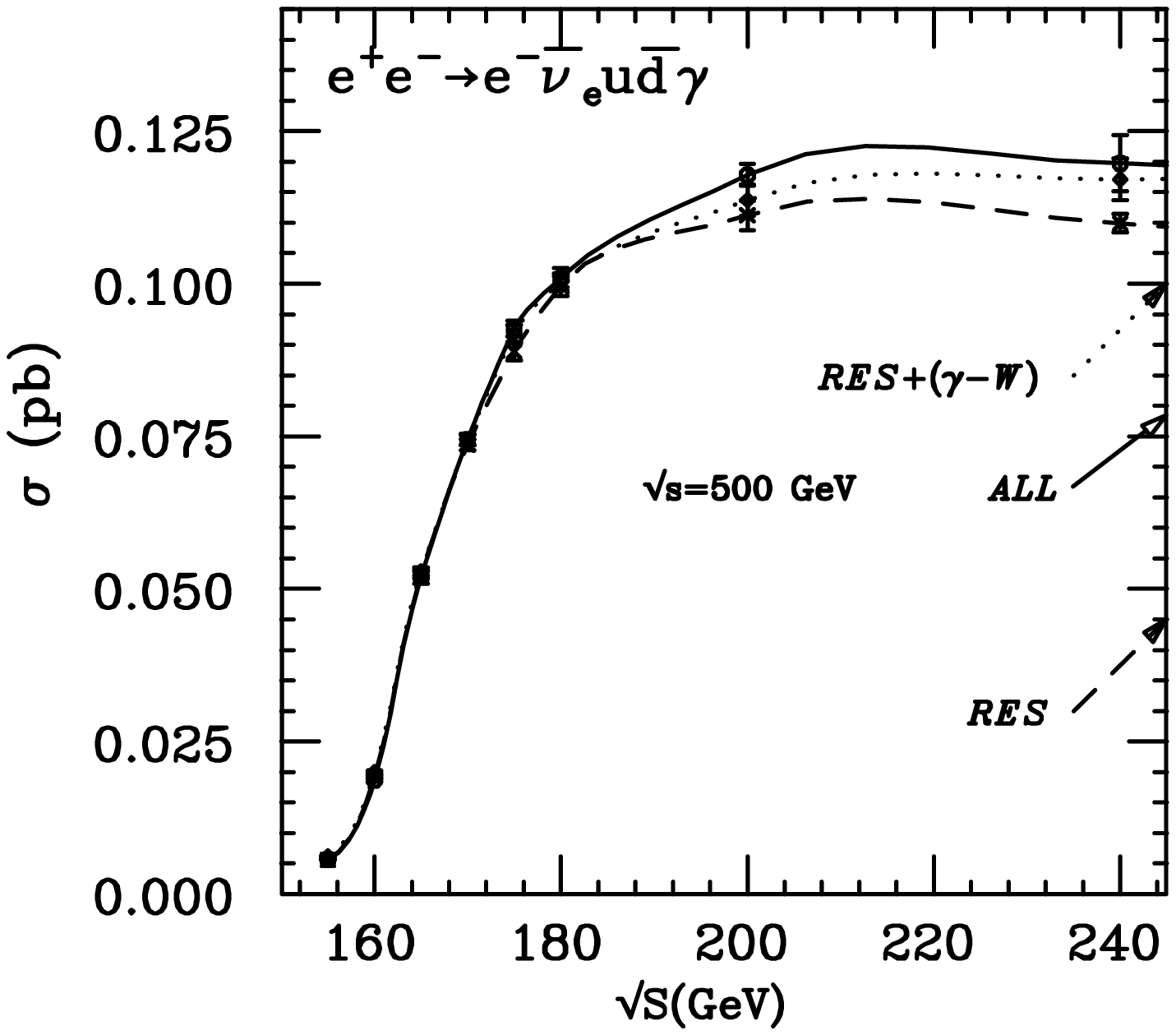,width=9.5cm,height=10.cm}}
%\vspace*{-1cm}
\vspace*{-1cm}
\caption{Total cross section for
$e^+e^- \to e^- \bar\nu_e u \bar{d}\gamma$
versus the C.M. energy with $\theta_e \ge 8^\circ$.
The solid line represents the complete
cross section while the dotted line the resonant term only.}
\vspace*{-0.5cm}
%\end{center}
\end{figure}
\subsection{The non-radiative case}
The contribution of the non-resonant diagrams has been considered for
two final states, $e^- \bar\nu_e u \bar{d}$ and
$\bar{u} d  u \bar{d}$. Fig.3 shows the cross sections versus
the center of mass energy around the W pair threshold. The solid line
represents the case where all diagrams are included,
the dashed line the contribution of the resonant diagrams only.
In the $\bar{u} d  u \bar{d}$ channel, one can observe
the onset of the Z resonant diagrams as a small shoulder starting
around $2M_Z$.
The behaviour below the W pair threshold can be better seen in fig.4
where the ratio of the three main resonant diagrams over the complete
calculation is plotted for the same range of center of mass energy.
In the
$e^- \bar\nu_e u \bar{d}$ channel, the effect is of the order of
$\pm3$\%  around
$\sqrt s$ = 160 GeV, while in $\bar{u} d  u \bar{d}$ it reaches $-$28\%.
At the energy of future LC, 500 GeV, the contribution of non-resonant
diagrams is quite significant in $e^-\bar\nu_e u\bar d$ but is
negligible in $\bar u d u \bar d$.

Similarly to the $\gamma-\gamma$ events, the non-resonant processes are
predominantly peaked in the forward/backward direction. Fig.5 shows
the behaviour of the cross section versus the electron polar angle cut.
The non-resonant contribution vanishes as the cut is raised.
\begin{figure}[ht]
%\begin{center
\vspace*{-1cm}
\mbox{\epsfig{file=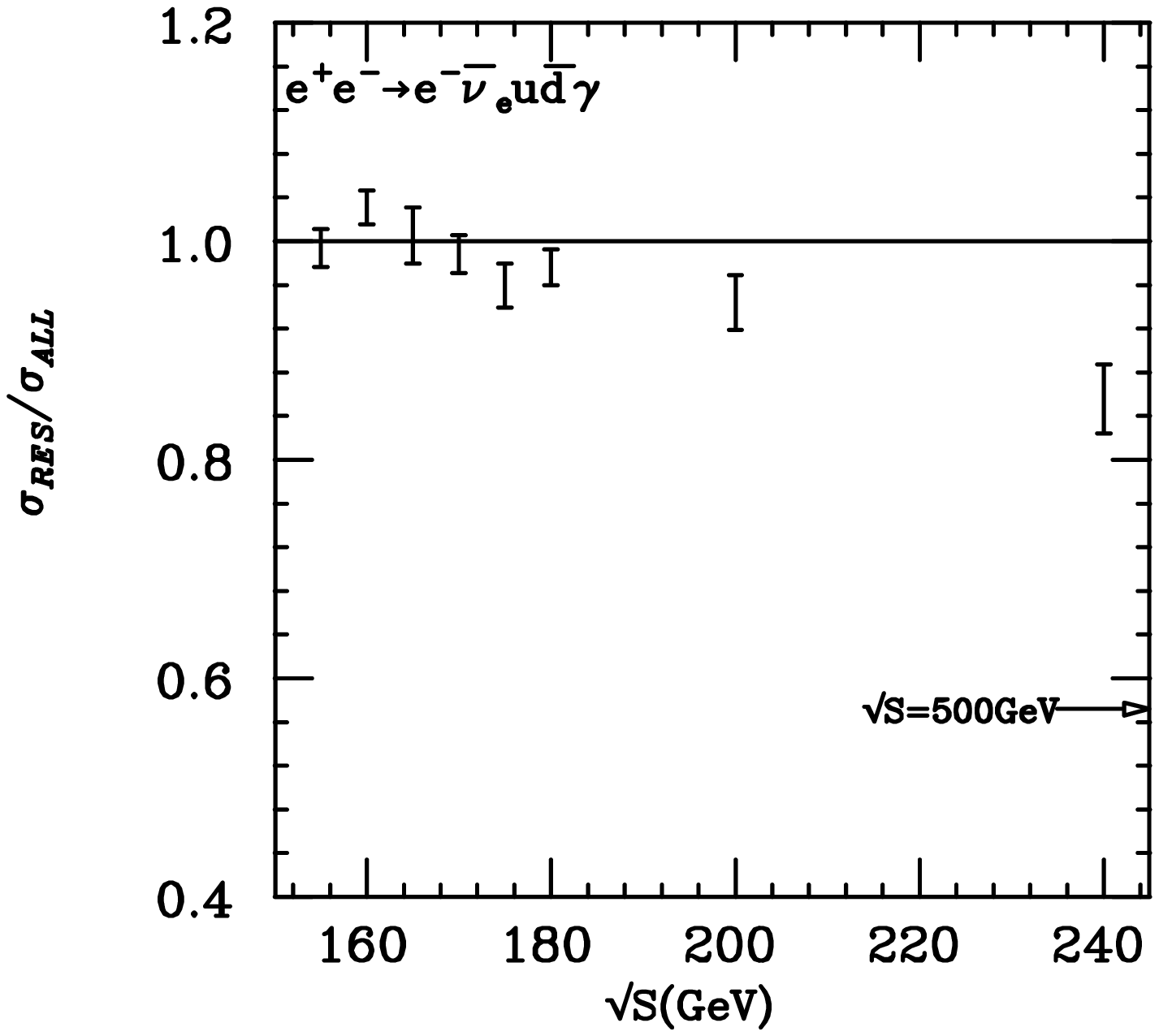,width=9.5cm,height=10.cm}}
\vspace*{-1cm}
\caption{Energy dependence of the contribution of non-resonant diagrams
in $e^+e^- \to e^-\bar\nu_e u \bar{d} \gamma$
with $\theta_e \ge 8^\circ$.
The ratio of contribution of
resonant diagrams versus all diagrams is shown.}
\vspace*{-0.5cm}
%\end{center}
\end{figure}
\begin{figure}[ht]
%\begin{center
\vspace*{-1cm}
\mbox{\epsfig{file=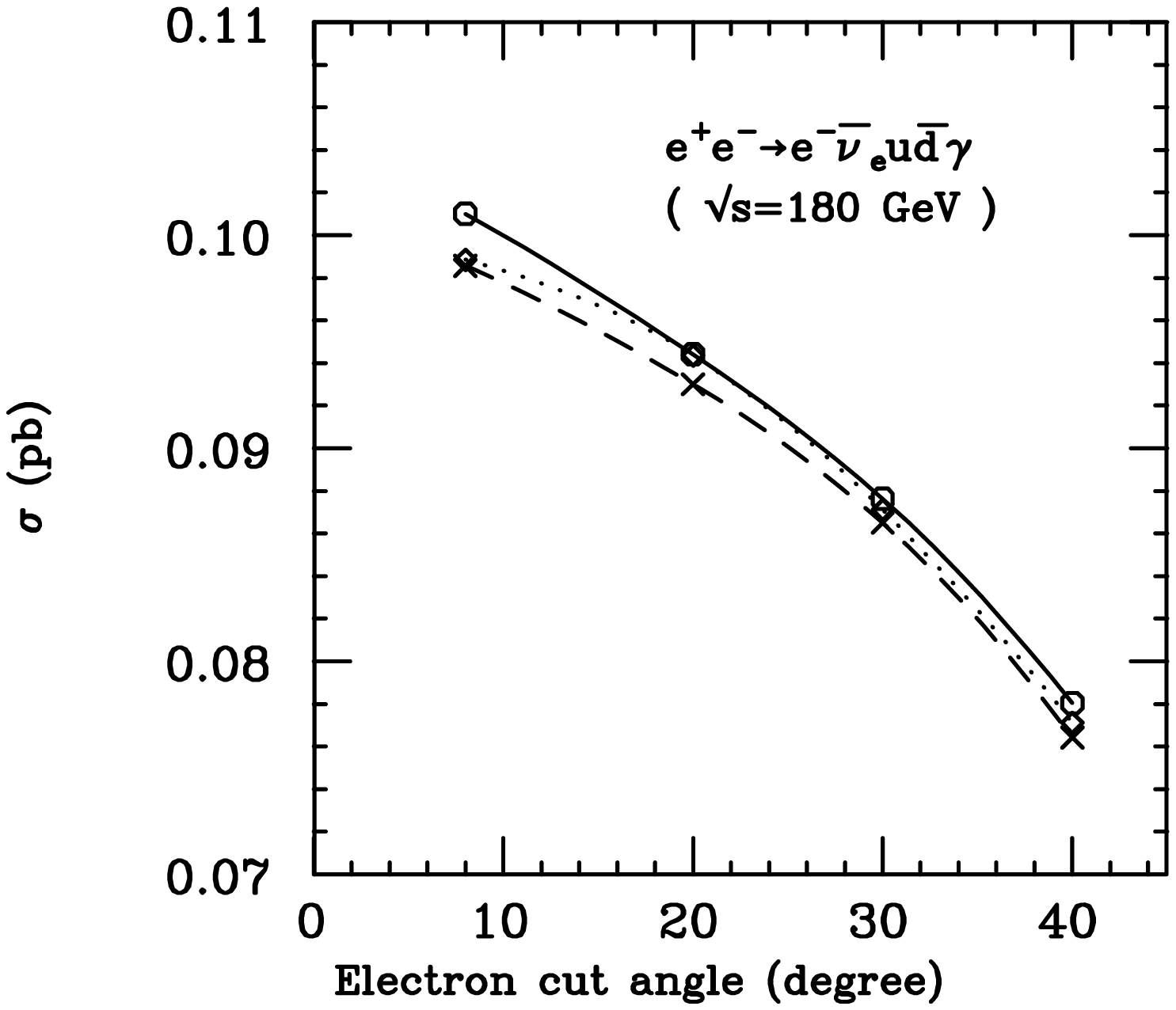,width=9.5cm,height=10.cm}}
\vspace*{-1cm}
\caption{Total cross section for $e^+e^- \to e^- \bar\nu_e u \bar{d}
\gamma$ versus the electron polar angle cuts in degree. The solid line:
all diagrams, the dashed line: resonant diagrams, and the dotted line:
resonant and $\gamma-W$.}
\vspace*{-0.8cm}
%\end{center}
\end{figure}
\subsection{The radiative case}
In the covariant gauge, the GRACE package generates 752 diagrams when
all the interactions are included, but actually 142 diagrams are used
in the unitary gauge with Higgs-fermion coupling being set to 0.
The full tree level computation requires, for a decent precision,
some 100 hours on a HP 735 computer. Thanks to the
vectorization of the integration package BASES, the most extensive
computation could be performed on the KEK vector processor within 1
hour.
%A similar analysis to the non-radiative case has been performed.
Fig.6 shows the contribution of the non-resonant terms in the total
cross section versus the C.M. energy. The behaviour is similar to
the non-radiative case. In fig.7,
one can see more clearly the contribution of
non-resonant diagrams similar to fig.4 for non-radiative case.

The dependence of the cross section on the electron polar angle cut
$\theta_e$ is presented in fig.8. Fig.9 shows the distribution of
the invariant mass of the pairs of final state fermions.
The general shape is, as expected, a Breit-Wigner distribution. However
some background coming from the onset of non-resonant processes can be
seen. In addition, for the radiative process (fig.10), a mass shift of
the order of 1 GeV is observed on the $e^- \bar\nu_e$
invariant mass due to final state radiation.
\begin{figure}[ht]
%\begin{center
\vspace*{-1cm}
\mbox{\epsfig{file=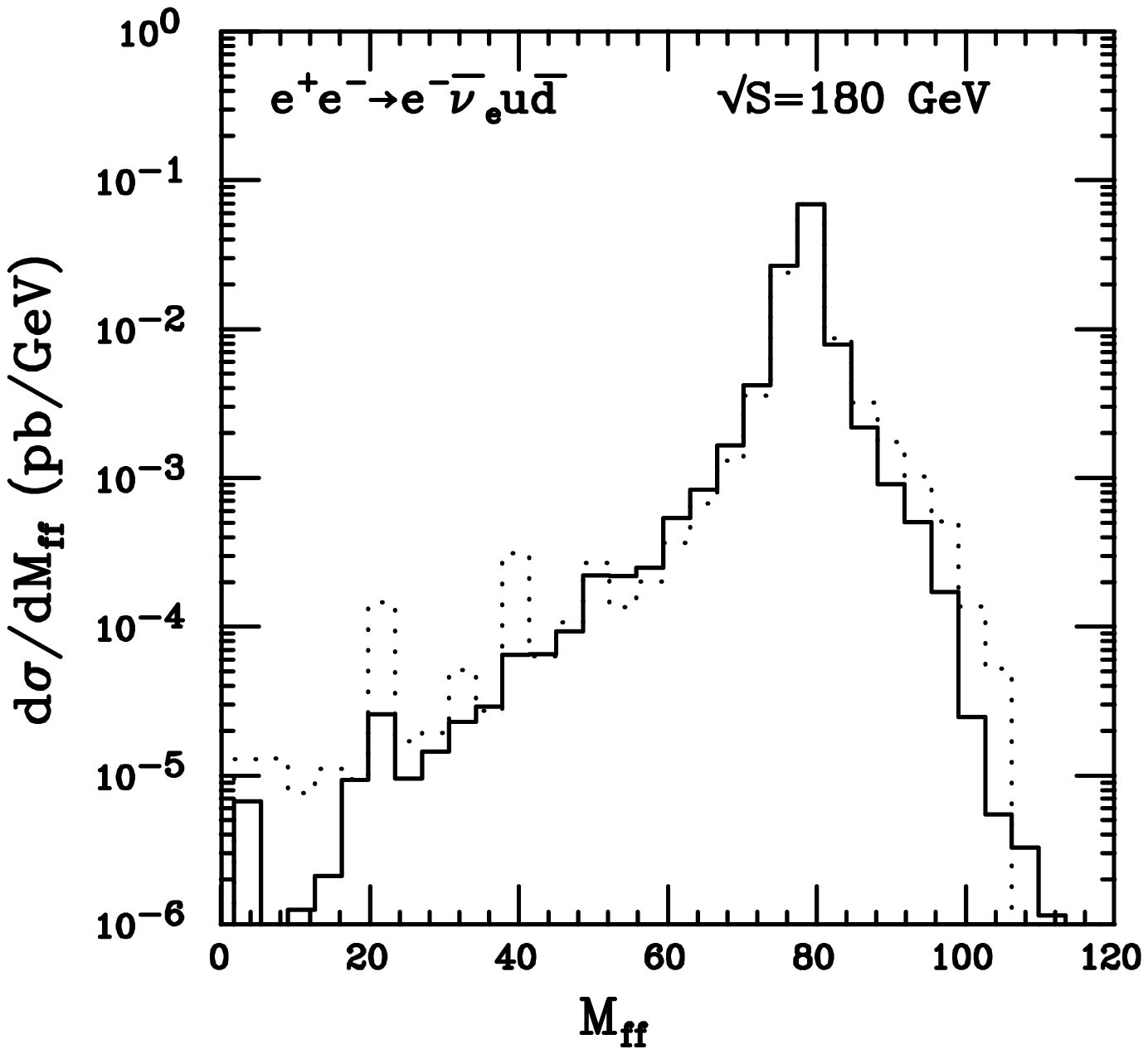,width=9.5cm,height=10.cm}}
\vspace*{-1cm}
\caption{Invariant mass for non-radiative process in solid line for
the $u \bar{d}$ system
and in dotted line for the
$e^- \bar\nu_e$ system
with $\theta_e \ge 8^\circ$.
}
\vspace*{-0.5cm}
%\end{center}
\end{figure}
\begin{figure}[ht]
%\begin{center
\vspace*{-1cm}
\mbox{\epsfig{file=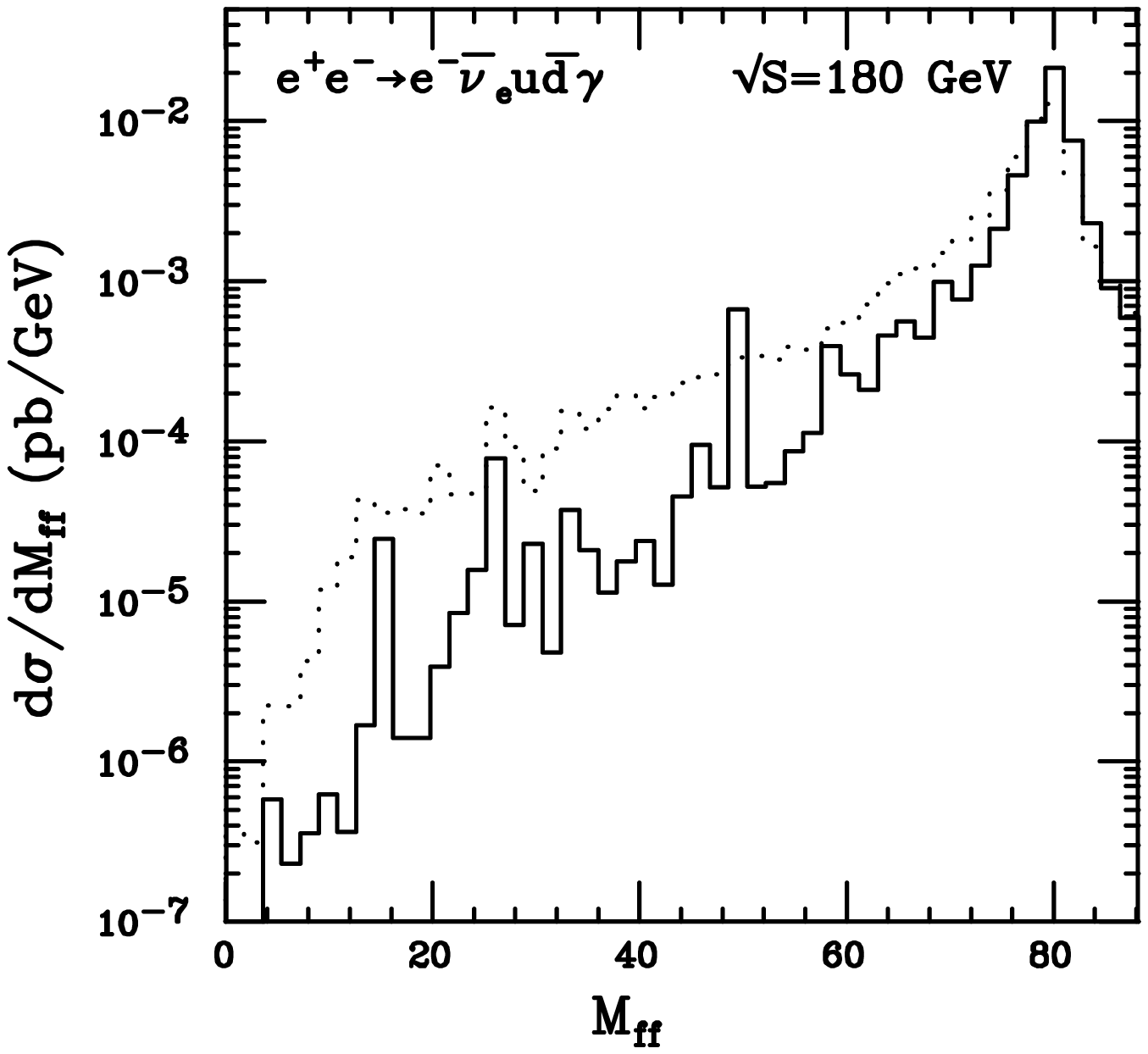,width=9.5cm,height=10.cm}}
\vspace*{-1cm}
\caption{Invariant mass for radiative process in solid line for
the $u \bar{d}$ system
and in dotted line for the
$e^- \bar\nu_e$ system
with $\theta_e \ge 8^\circ$.
}
\vspace*{-0.5cm}
%\end{center}
\end{figure}
\section{Conclusions}
A complete calculation of the radiative four-fermion final state
$e^+e^- \to e^- \bar\nu_e u \bar{d}\gamma $ has been performed,
for the first time, using automatic Feynman diagram
calculation techniques implemented in the GRACE system.
Comparison with two non-radiative computations
$e^-\bar\nu_e u \bar{d}$ and
$\bar{u} d u \bar{d}$ has been carried out.
The contribution of the non-resonant processes is large below threshold
(at least for $\bar{u} d u \bar{d}$ channel)
which makes the W mass determination from the excitation curve
more difficult.
At high energy, the contribution increases, reaching 46\%
($e^-\bar\nu_e u \bar{d}$)
and 42.5\% ($e^-\bar\nu_e u \bar{d} \gamma$) at $\sqrt s$ = 500 GeV.
Determination of the W mass from direct reconstruction
will have to
deal with the
intrinsic background from the non-resonant processes prominently
in the forward direction. Furthermore the final state radiative
correction will introduce a shift in the mass of the W unless the photon
is properly taken into account in the algorithm of mass
reconstruction. However, this last point may prove to be difficult
as the photon may be produced either from the initial
state, the intermediate charged boson or the final state fermions.
Only this last class plays a role in the W mass determination.
More detailed studies are needed to check if additional kinematical cuts
can improve the mass determination.

However to quantify the
actual relative contribution of radiative events in the total cross section,
soft photon and one-loop contributions must be
taken into account. Large negative contributions are expected, reducing,
consequently, the number of radiative events and softening the mass
reconstruction problem.
\section{Acknowledgements}
This work has taken place in the framework of the KEK-LAPP collaboration
supported in part
by the Ministry of Education, Science and Culture (Monbusho) under
the Grant-in-Aid for International Scientific Research Program
No. 04044158 in Japan and by le Centre National de la Recherche
Scientifique (CNRS) and l'Institut de Physique Nucleaire et Physique
des Particules (IN2P3) in France.

%\newpage


\begin{thebibliography}{99}
\bibitem{sotchi}
T. Ishikawa, T. Kaneko, S. Kawabata, Y. Kurihara, Y. Shimizu and
H. Tanaka, KEK Preprint 92-210, 1992 and in the {\sl Proceedings of
the 7th workshop on high energy physics and quantum field theory},
Sotchi, Russia, 1992.
\bibitem{pittau}
M. Pittau, in these proceedings.
\bibitem{veltman}
M. Lemoine and M. Veltman, {\sl Nucl. Phys.} {\bf B164 } (1980) 445
 \bibitem{shimizu}
H. Tanaka, T. Kaneko and Y. Shimizu, {\sl Comput. Phys. Commun.}
{\bf 64}(1991) 149.
\bibitem{other}
See related contributions in these proceedings.
\bibitem{grace}
T. Ishikawa, T. Kaneko, K. Kato, S. Kawabata, Y. Shimizu and H. Tanaka,
KEK Report 92-19, 1993, The Grace manual Ver. 1
\bibitem{chanel}
H. Tanaka, {\sl Comput. Phys. Commun.} {\bf 58} (1990) 153
and see also \cite{shimizu}.
\bibitem{bases}
S. Kawabata, {\sl Comput. Phys. Commun.} {\bf 41}(1986) 127.
\bibitem{boos}E.E. Boos {\sl et al.}, {\sl Phys. Lett.}
{\bf B326}(1994) 190.
\bibitem{pukhov}
A.E. Pukhov and V.A. Ilyin, private communication.
\end{thebibliography}
\end{document}